\documentclass[useAMS,usenatbib,usegraphicx]{mn2e}

\title{Radial velocities and binarity of southern SIM grid stars}
\author[Valeri V. Makarov \& Stephen C. Unwin]{Valeri V. Makarov$^{1}$\thanks{E-mail:
vvm@usno.navy.mil} \& Stephen C. Unwin$^{2}$\thanks{E-mail:
Stephen.C.Unwin@jpl.nasa.gov}\\\\
$^{1}$US Naval Observatory, 3450 Massachusetts Ave NW, Washington DC 20392-5420, USA\\
$^{2}$Jet Propulsion Laboratory, California Institute of Technology, Mail Stop 321-100, 4800 Oak Grove Drive, Pasadena, CA 91109, USA}
\begin{document}

\date{Accepted . Received ; in original form }

\pagerange{\pageref{firstpage}--\pageref{lastpage}} \pubyear{2002}

\maketitle

\label{firstpage}

\begin{abstract}
We present analysis of precision radial velocities (RV) of 1134 mostly red giant stars in the southern sky,
selected as candidate astrometric grid objects for the Space Interferometry Mission (SIM). Only a few 
(typically, 2 or 3) spectroscopic
observations per star have been collected, with the main goal of screening binary systems. The estimated
rate of spectroscopic binarity in this sample of red giants is 32\% at the 0.95 confidence level,
and 46\% at the 0.75 confidence. The true binarity rate is likely to be higher, because our method is not quite sensitive 
to very wide binaries and low-mass companions. The estimated lower and upper bounds of stellar RV jitter 
for the
entire sample are 24 and 51 m/s, respectively; the adopted mean value is 37 m/s. A few objects
of interest are identified with large variations of radial velocities, implying abnormally high mass ratios. 
 \end{abstract}


\section{Introduction}

Space Interferometry Mission (SIM) was designed to perform global and narrow-angle astrometry at an unprecedented 
precision level of 1-10 microarcseconds per single measurement. At the time of its termination by NASA (following
an explicitly negative assessment by the National Academy of Science in the Astro2010 Decadal Survey)
SIM was in an advanced Phase B, having passed all eight technology development milestones. One of the
ongoing Phase B studies was aiming at constructing a uniform, all-sky grid of reference stars, which would
serve as an important stepping stone toward generating a global reference frame at a $\simeq 4$ $\mu$as level
\citep{unw}. SIM would spend a considerable fraction of its operational life time observing some 1300 grid stars
in the wide-angle regime. The grid was needed to construct a rigid global reference frame from essentially
differential path delay measurements of objects within a $15\degr$-diameter field of regard \citep{mami}.
Establishing a set of verified, astrometrically stable grid stars was crucial for the success of the
global astrometry mission. In particular, a significant fraction of binary systems among the culled grid
objects could jeopardize the project.

Grid star candidates were selected by an elaborate system of criteria, deemed to maximize the rate of
stable, single stars. A detailed discussion of the criteria is rather technical and is outside the scope of this
paper. Candidate grid stars were selected among presumably early K and clump giants, with estimated distances between
500 and 1000 pc. Preference was given to candidates with small to moderate proper motions. The latter criterion was supposed to reduce the
risk of selecting nearby red dwarfs, since neither spectroscopic luminosity class nor parallaxes were available
for these 10--11 mag stars. The candidates were found mostly in the Tycho-2 catalog \citep{hog},
avoiding the objects listed in the Tycho Double Star Catalog \citep[TDS,][]{fab}, and from the Guide
Star catalog (GSC). Most of the sample stars have $B-V$ colors between 1.0 and 1.3 mag. All stars have declinations
south of $-21\degr$.

The SIM project office organized a large-scale observational campaign to screen spectroscopic binaries
in both celestial hemispheres. High-precision radial velocity (RV) observations of some 3500 grid candidates were to be
collected by three teams through a competitive contract on medium-class telescopes. One of the teams,
led by Swiss astronomers Didier Queloz and Damien Segransan, produced the largest and most precise set of
data spanning 753 days. The team used the 1.2 m Euler telescope in ESO La Silla Observatory, Chile,
equipped with the CORALIE spectrograph. The set-up of the instrument and the observing technique are similar to
those used for the exoplanet search program \citep{que}. Additional information on the calibration and data reduction techniques
can be found in \citep{bar}.

\section{RV, errors and binarity}
\begin{figure}
\includegraphics[width=90mm]{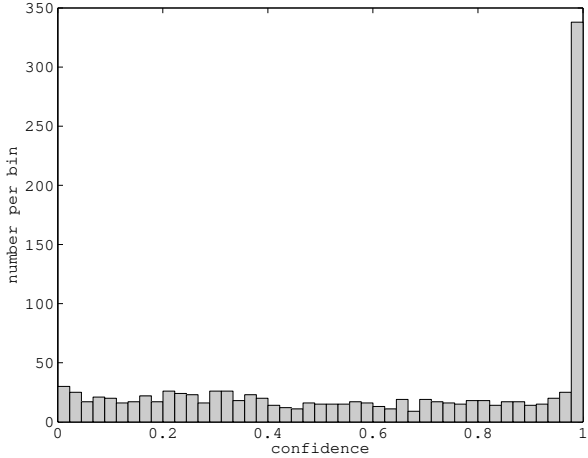}
\caption{Histogram of Spectroscopic binarity confidence levels for 1333 grid stars with more than 1 RV measurements.}
\label{pvalue.fig}
\end{figure}

The data set used in this paper is structured as shown in Table 1. A few lines of data are given for each of the
1134 observed stars. The first line is a header, consisting of the star's name, equatorial coordinates (right ascension and declination) on J2000, number of RV measurements $n_{\rm obs}$, 
derived mean RV in km s$^{-1}$ and
its uncertainty (including the estimated stellar jitter) in km s$^{-1}$, binarity confidence, binary flag (0 for
non-binary, 1 for binary) with 0.75 confidence, and binary flag with 0.95 confidence. The names are mostly
Tycho-2 identifications, which are consistent with the search engine of
the Simbad Database. The RV values given in the header are the weighted mean estimates. For binary systems,
these values may significantly deviate from the actual center-of-mass velocities. The heliocentric RV data are fully
calibrated for known instrumental effects, but they do not include any star-specific
corrections of astrophysical nature,
such as convective blueshift or gravitational redshift. Nor any corrections are implemented for a possible
difference between the observed and template K0III spectra. The header is followed by the specified number of
individual RV measurement lines. Each line contains the Julian date of observation, observed RV in
km s$^{-1}$, and the formal observational error in m s$^{-1}$. The latter is the estimated instrument error
only, and as such, it does not include any uncertainty associated with the physical nature of the object.
A total of 3026 observations are included in the table.

\begin{table*}
\caption{Radial velocity data for 1334 SIM grid stars.}
\label{rv.tab}
\begin{tabular}{@{}*{9}{l}}
\hline
7522-00774-1 & 359.965037 & -35.074871 & 4  & +41.3654  &  0.0603 & 0.905 & 1 & 0\\
\cline{1-3}
2453253.83186604  & +41.3602 & 14.3 &&&&&\\
2453556.93681788  & +41.4254 & 18.4 &&&&&\\
2453662.70871491  & +41.2784 & 23.5 &&&&&\\
2453989.76299457  & +41.3880 & 17.9 &&&&&\\
\hline
8021-00744-1 & 359.292000 & -44.504791 & 3  & +40.2173  &  0.0411 & 0.252 & 0 & 0\\
\cline{1-3}
2453369.56899134  & +40.2285 & 20.6 &&&&&\\
2453639.69474182  & +40.1933 & 14.1 &&&&&\\
2453986.70107572  & +40.2331 & 19.1 &&&&&\\
\hline
\end{tabular}
\\
\begin{flushleft}Notes:\\
The entire catalog is available online.\\
The columns in a header line are: 1) Star identification, mostly TYC, which are consistent with the Simbad search engine; 2) right ascension at J2000 in degrees; 3) declination at J2000 in degrees; 4) number of RV measurements; 5) mean radial velocity in km s$^{-1}$;
6) mean RV uncertainty including the estimated stellar jitter in km s$^{-1}$; 7) binarity confidence; 8) binary flag (0 for
non-binary, 1 for binary) with 0.75 confidence; and 8) binary flag with 0.95 confidence.
The columns in a data line are: 1) Julian date of observation; 2) observed RV in
km s$^{-1}$; and 3) the formal observational error in m s$^{-1}$.
\end{flushleft}
\end{table*}

Given only two or three measurements for most stars, not much can be done toward characterizing spectroscopic
binaries. We employed the simplest method of detecting possible binarity, i.e., a test of RV constancy.
The weighted mean RV is first computed from the available measurements. The residuals are then used to
compute the $\chi^2$ statistics. If our assumptions about individual random errors of RV measurements
are correct and there is no additional scatter due to binarity, the derived $\chi^2$ should follow the
$\chi^2$ distribution of probabilities with $k=n_{\rm obs}-1$ degrees of freedom. In particular, the
$\chi^2$ cumulative distribution function (CDF) specifies the probability of obtaining a value
which is smaller than the observed value. Reversely, $1-$CDF$(\chi^2)$ defines the p-value of
our statistical test, i.e., the probability of the null hypothesis to be rejected while being true.
The null hypothesis is that the star is single, and the observed scatter is solely caused by random
errors with the expected variance. Thus, the derived CDF$(\chi^2)$ can serve as the confidence of
binarity detection. 

The main difficulty in this application is that the estimated formal errors of RV measurements only
capture the instrumental part of the error budget, mostly the photon shot noise. A significant part,
and in fact, the larger part of the observed dispersion in RV for giants, comes from the astrophysical
phenomena on the stars themselves. This component, often called RV jitter, may be caused by stochastic
and uncorrelated appearance of photospheric features on the surface, such as bright plages or dark
spots, similar to the structures observed on the Sun. The changes in the distribution of surface brightness
result in correlated photometric and RV variations. The latter have a more complex character, being
a convolution of surface brightness distribution with the differential rotation curve and the geometric
projection onto the viewing direction. The basic relations for the photometric and RV jitter due to
spots are given in \citep{mak}. Most of the parameters entering these relations are not known for
field red giants, such as rotation velocities, typical relative surface area of spots, inclinations
of the equator to the line of sight. Little can be said about photospheric jitter on theoretical grounds.
If the magnetic activity features are long-lived and large, such as observed on most magnetically active
stars, many of these parameters can be reasonably estimated by intricate inverse solutions \citep{wal},
but this is not valid for regular red giants. Non-radial pulsations in red giants with periods between 1 hour and 3 days
with photometric relative amplitudes peaking at 400 ppm
may provide another significant contribution to the observed astrophysical jitter, but this assumption can only
be verified by simultaneous, high-accuracy RV and photometric measurements at high cadence.

If we knew which of the observed star are spectroscopically single, we could easily estimate the sample-mean
jitter. The mean jitter is quadratically added to the instrumental error for each observation, and the median
$\chi^2$ over the sample of single stars should be close to the argument of CDF at 0.5. In other words,
half of the single stars should have p-values smaller than 0.5. But it is not known a priori which stars are
single, therefore, we can estimate only robust bounds for the jitter component. The upper bound is derived
from the assumption that the observed scatter is only due to observational error and jitter, i.e., all the stars in
the sample are non-binary. The level of jitter is adjusted until the median $\chi^2$ is equal to the expected
value $k(1-2/(9k))^3$ \citep{wil}. 
The upper bound for the mean jitter thus estimated is 51 m s$^{-1}$. The actual value is certainly smaller, because
a significant fraction of the sample are strongly perturbed by binarity. The lower bound for jitter comes from
the assumption that only the higher half of sample values $\chi^2$ are affected by spectroscopic binarity.
The jitter level is adjusted until the median $\chi^2$ of the lower half of the sample distribution reaches
the expected value $k(1-2/(9k))^3$. The lower bound comes up to 24 m s$^{-1}$. We adopt the mean of the two
bounds, 37 m  s$^{-1}$, as the average jitter dispersion for the sample. This value is likely to be slightly
overestimated because, as we will see shortly, the rate of detectable binaries seems to be closer to 0.5
than to 0.

With this amount of jitter quadratically added to the formal instrument errors, the resulting distribution
of confidence levels is shown in Fig. \ref{pvalue.fig}. We note that the distribution of confidence levels is quite flat everywhere
except the highest bin. The flatness of this distribution is a significant result in itself, pertaining to the
nature of binary systems with red giant primaries. With 2 degrees of freedom, the $\chi^2$ reaches 10 for the corresponding CDF value
of $\simeq 0.993$. The mass of companion in a 1 year orbit corresponding to this detectable signal is of the order of $0.005\,M_{\sun}$.
In principle, even giant exoplanets can be detected with 3 observations.
The range of confidence levels between 0.993 and 0.99999 corresponds to the range 0.005 - $0.008\,M_{\sun}$ of companion mass,
thus, these data should be sensitive to super-Jupiters and brown dwarf companions in 1-year orbits. The absence of significant excess
at these confidence levels seems to confirm the ``brown dwarf desert" phenomenon, i.e., the paucity of substellar-mass
secondary companions in binary systems. However, when the jitter is set at the lower bound level of
24 m  s$^{-1}$, we detect a small upturn of the observed $CDF(\chi^2)$ distribution starting already at 0.95.
This CDF value corresponds,
very approximately, to companion masses of $0.003\,M_{\sun}$ for both 1 and 2 degrees of freedom.
Hence, these results do not preclude the existence of a limited number ($\sim20$) of super-Jupiter planets or brown dwarf
companions in our sample, which can be found among the marginally significant positives.

At a threshold confidence level of 0.95, the rate of spectroscopic binaries is 32\%. The binarity rate comes up to
46\% for the lower confidence threshold 0.75. As previously discussed, our accepted jitter value (37 m  s$^{-1}$)
may be overestimated. The true binarity rate is likely between 37 and 46 \%. If we set the jitter
component at the lower bound of 24 m  s$^{-1}$, the estimated rate of binaries comes up to 41\% at 0.95 confidence.
This number may include the contribution of substellar mass companions.

\section{Objects of note}
A few stars in the sample have unusually large RV variation, implying exceptionally high mass ratios or very short
orbital periods, or both. We select two such objects, which deserve a careful follow-up investigation.

{\bf TYC 6948-00350-1 = HD 206092} was observed with a total RV range of 266 km s$^{-1}$. This G9 III star has been
identified by \citet{kir} as rotationally variable with an astonishing period of $2.1876$ d and a line broadening
parameter $v\sin\,i$ of 80 km s$^{-1}$. Now it is also detected as a spectroscopic binary with a RV semi-amplitude
of at least 133 km s$^{-1}$. Assuming that the primary star is synchronized, and the rotation period equals the orbital
period, the mass ratio for this system should be above 1, or the eccentricity should be high. The latter option is
not feasible in the light of previous studies of rotational velocities of red giants in binary systems \citep{mas},
which found that all binaries with periods shorter than 20 d are circularized. We attempted to fold the available
RV measurements with the rotational period, but failed to obtain a clear phase curve. Follow-up photometric and spectroscopic
observations are needed to reveal the nature of this enigmatic object.

{\bf TYC 7381-00433-1 = HD 318347} has a RV range of at least 212 km s$^{-1}$. Little is known about this star.
Surprisingly though, it is listed in Simbad as O$^+$ type with emission lines, that is, a very hot and massive star.
However, it is quite red with a $B-V=0.92$ determined by \citet{dri}. We surmise this binary star may include a regular
red giant accompanied by a post-AGB star or some other rare type of object on a short-period orbit.

\section{Discussion}
Our estimate of multiplicity fraction for 1333 randomly selected, field red giant stars is close to the
more accurately known rate for nearby Solar-type stars, 46\%$\pm$2\% \citep{rag}. Our best estimate is 41\%, which may
or may not include companions of substellar mass, depending on the actual level of stellar RV jitter and its distribution.
The method of binarity detection employed in this paper is not sensitive to binaries with orbital periods longer than several years, possibly lowering this estimate by several percent. The wider-separation pairs are to be detected by direct imaging
\citep{tok} or astrometric techniques \citep{makzh}. They may contribute a few more percent to the all-inclusive multiplicity
fraction, although the wide and common proper motion companions are often found in hierarchical multiple systems \citep{make}.
On the other hand, dwarfs more massive than the Sun, which are more likely to be the progenitors of the present-day K
giants, have a slightly higher binarity rate than their sub-solar counterparts \citep{duc}. The overarching conclusion is
that there seems to be no significant difference in the incidence of binarity among main-sequence dwarfs and field red giants.

Our results are consistent with other observational data for field red giants. \citet{set} collected and analyzed RV data for
77 G and K giants and detected a 20\% multiplicity rate, despite the preliminary binarity screening. The RV jitter for
those bluer and less luminous giants not detected as binary was constrained to $<60$ m  s$^{-1}$. Based on the much larger
sample and more observations, we are able to drastically improve these estimates.
\section*{Acknowledgments}
Part of the research described in this paper was carried out at the Jet Propulsion Laboratory, California Institute of Technology, under contract with the National Aeronautics and Space Administration.

\label{lastpage}


\begin{thebibliography}{99}
\bibitem[\protect\citeauthoryear{Baranne et al.}{1996}]{bar} Baranne A., et al. 1996, A\&AS, 119, 373
\bibitem[\protect\citeauthoryear{Drilling}{1991}]{dri} Drilling J.S. 1991, ApJS, 76, 1033
\bibitem[\protect\citeauthoryear{Duch{\^e}ne \& Kraus}{2000}]{duc} Duch{\^e}ne G., Kraus A. 2013, ARA\&A, 51, 269
\bibitem[\protect\citeauthoryear{Fabricius et al.}{2002}]{fab} Fabricius C., et al. 2002, A\&A, 384, 180
\bibitem[\protect\citeauthoryear{H{\o}g et al.}{2008}]{hog} H{\o}g E., et al. 2000, A\&A, 355, 27
\bibitem[\protect\citeauthoryear{Kiraga}{2012}]{kir} Kiraga M. 2012, Acta Astronomica, 62, 67
\bibitem[\protect\citeauthoryear{Makarov \& Milman}{2005}]{mami} Makarov V.V., Milman, M. 2005, PASP, 117, 757
\bibitem[\protect\citeauthoryear{Makarov, Zacharias \& Hennessy}{2008}]{makzh} Makarov V.V., Zacharias N., Hennessy G.S.
 2008, ApJ, 687, 566
\bibitem[\protect\citeauthoryear{Makarov et al.}{2009}]{mak} Makarov V.V., et al. 2009, ApJ, 707, L73
\bibitem[\protect\citeauthoryear{Makarov \& Eggleton}{2009}]{make} Makarov V.V., Eggleton, P.P. 2009, ApJ, 703, 1760
\bibitem[\protect\citeauthoryear{Massarotti et al.}{2008}]{mas} Massarotti A., et al. 2008, AJ, 135, 209
\bibitem[\protect\citeauthoryear{Queloz et al.}{2000}]{que} Queloz D., et al. 2000, A\&A, 354, 99
\bibitem[\protect\citeauthoryear{Raghavan et al.}{2010}]{rag} Raghavan D., et al. 2010, ApJS, 190, 1
\bibitem[\protect\citeauthoryear{Setiawan et al.}{2004}]{set} Setiawan J., et al. 2004, A\&A, 421, 241
\bibitem[\protect\citeauthoryear{Tokovinin et al.}{2012}]{tok} Tokovinin A., et al. 2012, AJ, 144, 7
\bibitem[\protect\citeauthoryear{Unwin et al.}{2006}]{unw} Unwin S.C., et al. 2008, PASP, 120, 38
\bibitem[\protect\citeauthoryear{Walker et al.}{2007}]{wal} Walker G.A.H., et al. 2007, ApJ, 659, 1611
\bibitem[\protect\citeauthoryear{Wilson \& Hilferty}{1931}]{wil} Wilson E.B., Hilferty M.M. 1931, Proc. Natl. Acad. Sci. USA, 17, 684

\end{thebibliography}
\end{document}